\documentclass{PoS}
\newcommand{\tw}{\textwidth}
\newcommand{\ig}{\includegraphics}

\title{Excited state baryon spectroscopy from lattice QCD}

\ShortTitle{Excited state baryon spectroscopy from lattice QCD}

\author{\speaker{Stephen Wallace}  
       \thanks{\bf for the Hadron Spectrum Collaboration.}\\
        Department of Physics, University of Maryland, College Park, MD 20742, USA\\
        E-mail: \email{stevewal@physics.umd.edu}}
 
\abstract{
Lattice QCD calculations are presented for the spectra of $N^*$ excited states with spins
up to $J = \frac{7}{2}$.  Ambiguities of the standard method of spin identification are shown to be overcome by the use of lattice operators that transform according to $SU(2)$ symmetry restricted to 
the lattice.  Such operators are labeled by their continuum spins. Overlaps of the operators 
with the states obtained by diagonalizing matrices of correlation functions provide a clear 
link between continuum spins and lattice states, allowing spins to be identified.  Evidence 
for an approximate realization of rotational symmetry in the $N^*$ spectrum is presented.
In simulations with $m_{\pi} \ge  392$ MeV, the 
low-lying excited states of lattice QCD are found to have the same quantum numbers as 
the states of $SU(6)\otimes O(3)$ symmetry.  The lattice spectra are inconsistent with either
a quark-diquark model or parity doubling of states and they suggest that the $J = \frac{1}{2}^+$ 
Roper resonance may 
have a complex structure consisting of contributions from $L=0,\ 1$ and $2$.     
}

\FullConference{XXIX International Symposium on Lattice Field Theory \\
                 July 10 - 16 2011\\
                 Squaw Valley, Lake Tahoe, California}

\begin{document}

%%%%%%%%%%%%%%%%%%%%
\section{Introduction}
%%%%%%%%%%%%%%%%%%%%
 
       Substantial progress has been made in
       recent years by using lattice methods to solve QCD and, thus, to calculate the spectrum of 
       baryonic and mesonic excited states from first principles.  
%       However,  much additional work is planned in order to arrive at final answers
%        for the spectra of hadronic masses.  For example, the mass of the $\pi$-meson used in this report is 
 %      $m_{\pi} =$ 396 MeV and it ultimately will be lowered to the physical value of 135 MeV in order to 
  %     be realistic.  The lattice volume used is about $(2 fm)^3$ and calculations with larger volumes 
  %     will be performed to ensure that lattice states, particularly multiparticle states, are not ``squeezed'' by being in too small a box.  
  %     The spatial lattice spacing used is $a_s = 0.123$ fm.  Variations of spectra with decreasing lattice 
  %     spacings will be determined in order to allow an  extrapolation to the continuum limit,
   %    $a_s \rightarrow 0$.  Other improvements also are being addressed.  One is the inclusion 
   %    of multiparticle operators, such as $\pi N$,  in order to provide robust access to the 
   %    decay channels for baryonic excited states, which should be treated as 
    %    resonances.   Another improvement is the identification of spins in lattice calculations.    
       
       In this report, we describe a recent solution to the problem of spin
        identification for baryonic excited states.  This development has made it possible 
        to determine patterns in the lattice spectra of $N$ and $\Delta$ excited states 
        that have the same spins and numbers of states as in
        $SU(6)\otimes O(3)$ symmetry.  
        The lattices and methods used to identify spins are described fully in Ref.~\cite{Edwards:2011} and the reader
      should consult that paper for details about the calculations.  %After a 
    %    brief description of some key features of the operators used, this review focuses on 
    %  the insights that follow from the determination of baryon spins in lattice calculations. 

The standard method of spin identification relies on patterns of degenerate states across 
irreps of the octahedral group in the continuum limit.  For
half-integer spins one uses the double-covered octahedral group.
This method has been found to be ambiguous for two simple reasons. First, there is a high degree of degeneracy in lattice spectra and second, the energies of states are subject to uncertainties owing to fluctuations of the gauge configurations.  A typical consequence is that approximately degenerate baryon states in $G_1$, $H$ and $G_2$ irreps could indicate a
spin $\frac{7}{2}$ state, but it could equally well indicate an accidental degeneracy of a 
spin $\frac{1}{2}$ state and a spin $\frac{5}{2}$ state.  This ambiguous situation is not likely to improve
without a new method.   
 
       The key developments that have enabled spin identification 
       and, thus, better insight into
       the lattice spectra are i.) the construction of lattice operators 
       that transform as irreducible 
       representations (irreps) of the $SU(2)$ rotational symmetry restricted to the lattice, and 
       ii.) the use of operator overlaps to tell which operators create which states.  
       
       The operators are labeled with
       a continuum spin quantum number, $J$.  They are realized by 
       incorporating combinations of covariant lattice derivatives that transform as
       orbital angular momenta restricted to the lattice.  Combinations of the baryon 
       spin, S, and orbital angular 
       momenta, L $\leq$ 2, are constructed to provide total spins $J \leq \frac{7}{2}$.  A similar
       development for integer spins (meson states) was given 
       in Ref.~\cite{Dudek:2009}. 
        
       In order to be suitable for lattice calculations,   
       operators labeled by continuum spin $J$ must be subduced to lattice operators that transform as irreps of the 
       octahedral group.  Because of the symmetries of a cubic lattice, it is the latter operators that 
       provide an orthogonal basis for calculations.  Subduction is performed using matrices that
       provide the change of basis from $SU(2)$ quantum states, labeled as $|J,M\rangle$,
       to quantum states that are irreps of the octahedral group in the continuum limit labeled as $|\Lambda, r,[J]\rangle$, where 
       $\Lambda$ and $r$ denote the irrep and row while $[J]$ denotes the continuum spin.  The
        matrix elements $\langle J,M | \Lambda,r,[J]\rangle$ provide the subduction matrices. 
       
       %%%%%%%%%%%%%%%%%%%%%%%%%%%%%%%%%%%%%%%%%%%%
       \section{Approximate rotational symmetry and spin identification}
       %%%%%%%%%%%%%%%%%%%%%%%%%%%%%%%%%%%%%%%%%%%%
\begin{figure*}
 \hspace{1in} \includegraphics[height=.45\textheight]{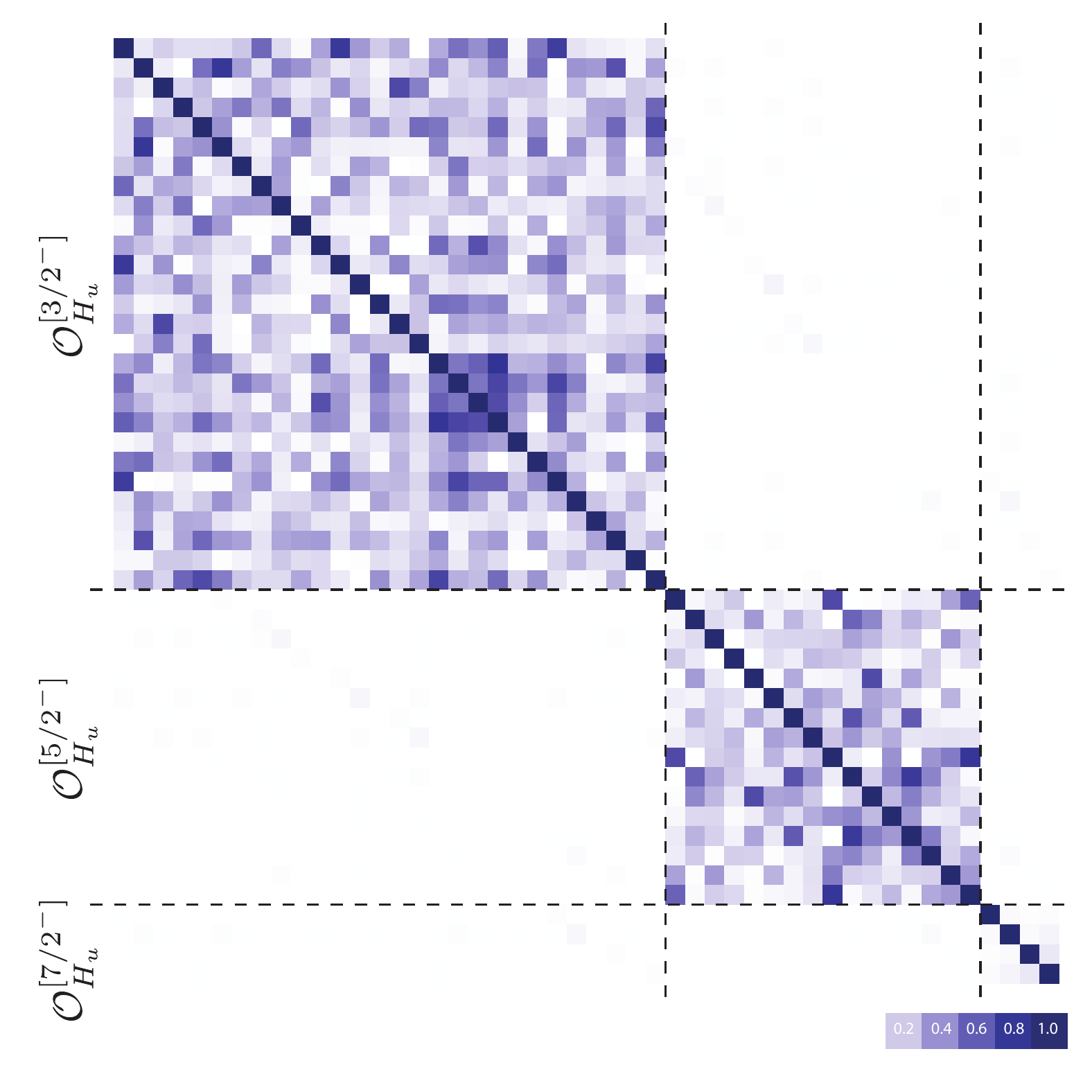}
  \caption{Matrix of $H_u$ correlation functions at time slice 5. \label{fig:J_blockdiag} }
\end{figure*}
       
       Even though the lattice spacing used in this work 
       is not especially small, good evidence is found for an approximate realization of  the continuum
       rotational symmetry in the lattice spectra.  An example of this is given in 
       Fig.~\ref{fig:J_blockdiag}, which indicates
       the relative sizes of matrix elements of a $48\times 48$ matrix of nucleon correlation functions 
       in the $H_u$ irrep of the double-covered octahedral group. The $u$ subscript of $H_u$ denotes 
       ungerade, or negative parity.  Each pixel indicates the magnitude
       of one matrix element according to the scale given in the lower right corner. The 48 operators 
       used consist of 28 operators that are 
       subduced from $J=\frac{3}{2}$, whose matrix elements are in the upper block; 16 
       operators that are subduced from $J=\frac{5}{2}$, whose matrix elements are in the 
       middle block; and 4 operators that are subduced from $J=\frac{7}{2}$ in the lower block.  
       Matrix elements that are outside the blocks involve operators subduced from different $J$ values.
       They are all quite small.  Thus, the matrix is close to
       block diagonal in spin, $J$, which is a signal of the approximate realization of rotational 
       symmetry.  
       
       Approximate rotational symmetry also is clearly evident in spectra that are obtained
       from matrices of correlation functions.  When the matrices are diagonalized, they may
       be expressed (at large time t) as a spectral decomposition,
\begin{equation}
        C_{ij}(t) = \sum_n \frac{Z_i^{n*} Z_j^{n}}{2m_n} e^{-m_n t},
        \label{spectro_decomp}
\end{equation}
    where $i$ and $j$ are labels of the operators ${\cal O}_i$ and ${\cal O}_j$ that are involved in the correlation function, the sum is over eigenstates with $m_n$ being the mass of the $n^{th}$ eigenstate and 
      ``overlap factors", $Z^n_i \equiv \langle n | {\cal O}_i^\dag | 0 \rangle$ 
       are matrix elements for the $i^{th}$ operator to create the
       $n^{th}$ eigenstate.
       The masses $m_n$ determine the spectrum and the overlap factors show which operators create which states.  
     
          A detailed examination of the overlap factors shows that lattice states are created 
          predominantly by operators subduced from a
       single continuum spin, $J$.  We identify the spin of each lattice state as the 
       continuum spin of those operators.

%%%%%%%%%%%%  
\begin{figure*}
\hspace{0.6in}\includegraphics[height=0.58\tw,width=0.7\tw,bb=20 30 420 396]{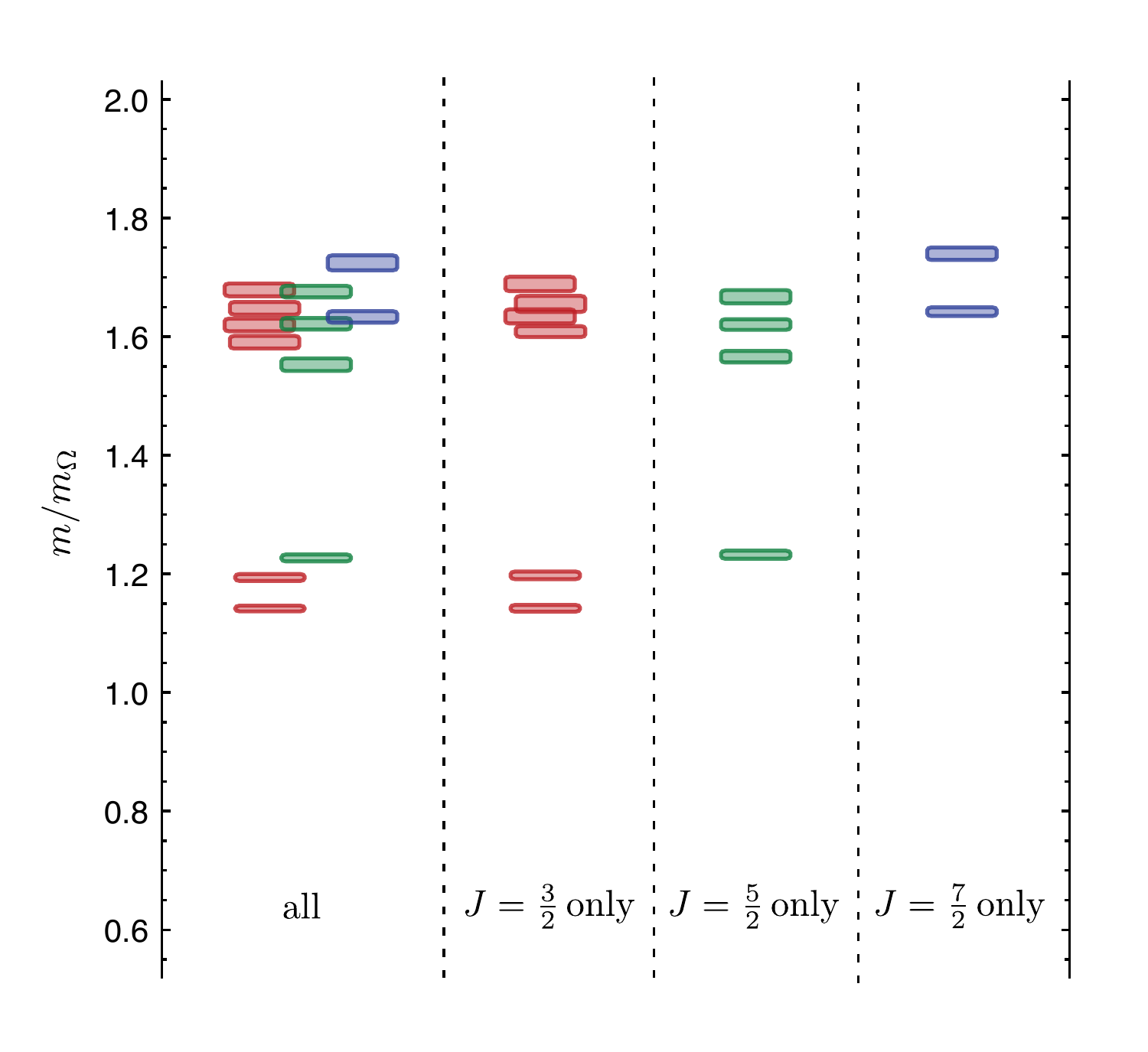}
\caption{Extracted Nucleon $H_u$ mass spectrum for various operator bases. Column 1 uses all
48 $H_u$  operators, columns 2, 3 and 4 use restricted bases of 28 $J=\frac{3}{2}$ operators, 
16 $J=\frac{5}{2}$ operators and 4 $J=\frac{7}{2}$ operators, respectively.  States with spin 
identified as $J = \frac{3}{2}$ are shown in red, $J = \frac{5}{2}$ are shown in green  and 
$J = \frac{7}{2}$ are shown in blue. Results are from the $m_{\pi} $ = 524 MeV ensemble. }

\label{fig:vary_ops}
\end{figure*}

Once spins have been identified, a further test of the approximate rotational invariance 
in the lattice spectra becomes possible.
If indeed the couplings are small between states subduced from different continuum spins, 
then the omission of such couplings should not much affect the excited state spectra. 
That proposition can be tested  by calculating energies using all operators, 
and comparing them with the energies obtained from the subset of operators subduced 
from a single $J$ value.
If approximate rotational invariance is achieved in the spectrum, the 
energies should be nearly the same.  In Fig.~\ref{fig:vary_ops}, we show such a comparison 
for the states in the Nucleon $H_u$ irrep.  The left column of Fig.~\ref{fig:vary_ops}, 
labelled ``all'', shows the lowest 12 energy levels obtained from matrices 
of correlation functions using the set of all 48 $H_u$ operators, with spins identified using the 
overlaps, as described above. The second column shows the lowest 6 levels resulting from the variational method when the operators are subduced from continuum 
spin $J=\frac{3}{2}$.
Similarly, the third column shows the lowest 4 levels obtained from the operators subduced from continuum spin $J=\frac{5}{2}$, and the last column shows the lowest two levels from operators subduced  from $J=\frac{7}{2}$.

The results are striking. 
We see that the masses of the levels in each of the restricted bases agree quite well with the results
found in the full basis. 
The agreement is quite remarkable because one expects that
operators in the $H_u$ irrep that can couple with the ground state (the lowest 
$J=\frac{3}{2}$ state), will show a rapid decay of correlators down to the ground state as
a function of time, $t$.  However, the 
higher-energy spin $\frac{5}{2}$ and $\frac{7}{2}$ states do not show such a 
decay; we obtain good fits to correlators at large $t$, where a single mass dominates the decay.  
These results provide a rather striking demonstration for the lack of significant rotational symmetry
breaking in the spectrum.

%%%%%%%%%%%%%%%%%       
    \section{Lattice $N^*$ spectrum with spins identified}
%%%%%%%%%%%%%%%%%    

\begin{figure*}
\vspace{-.3in}\hspace{0.5in}\ig[width=0.8\tw]{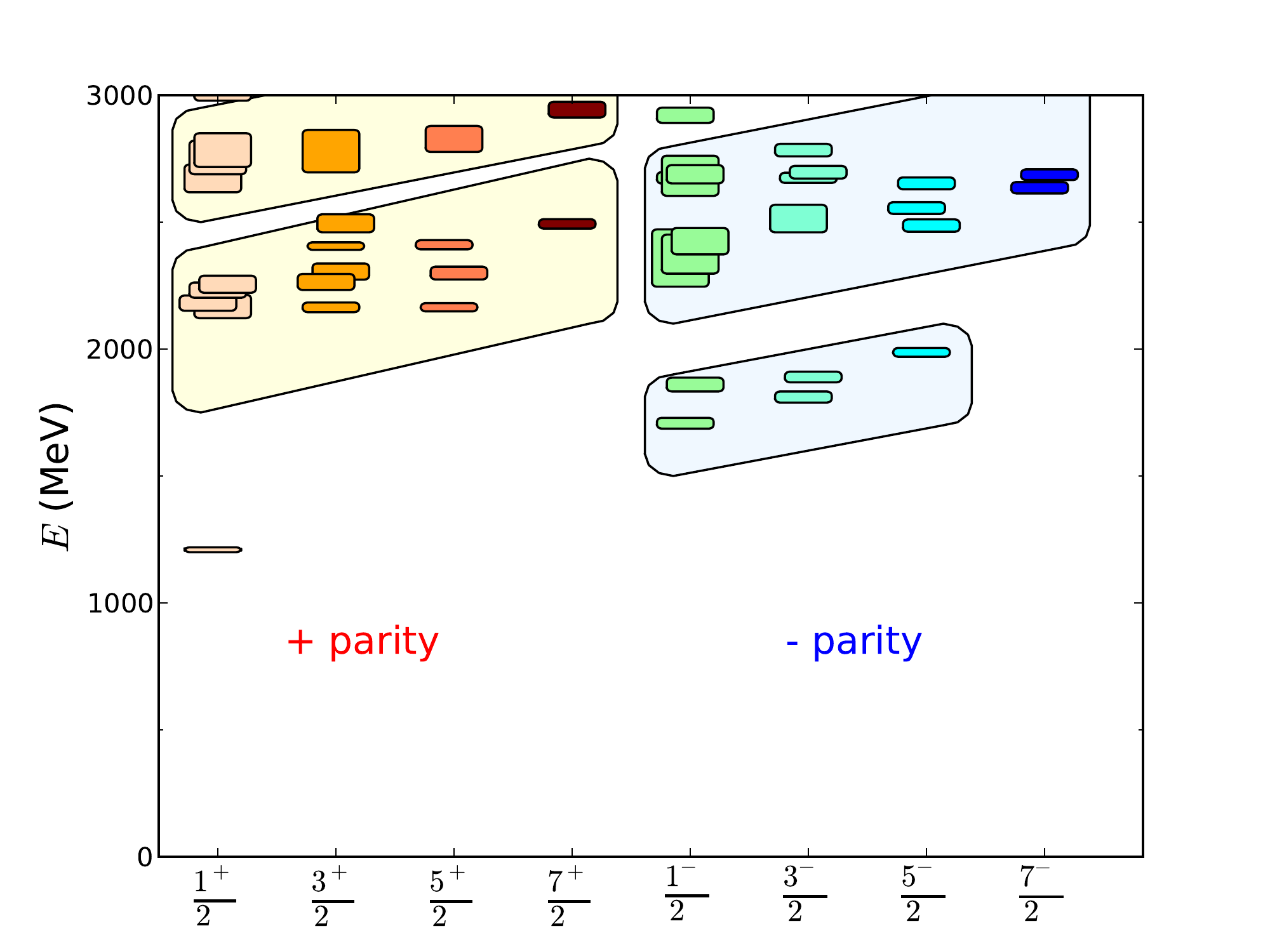}
\caption{Lattice $N^*$ energy spectrum in columns of good $J^P$ at $m_{\pi}$ = 396 MeV with spins identified.  Shaded regions 
show bands of states with alternating parity. \label{fig:J_spectrum_bands}}
 \end{figure*}

        The robust identification of spins allows the 
       spectrum to be displayed as in Fig.~\ref{fig:J_spectrum_bands}, with masses and uncertainties 
       indicated by boxes in columns labeled by the spin and parity values of the states, $J^P$.     
       The excited nucleon spectrum of Fig.~\ref{fig:J_spectrum_bands} contains many more states than 
       the experimental spectrum.  All the lattice states, even multiparticle states, are discrete levels (within 
       statistical uncertainties) because of the use of a finite box. 
      Moreover, the use of three-quark operators, as in this work, does not give significant multiparticle 
      contributions, although they should be present in principle.  Thus, the expected washing out of highly excited states 
      when they are embedded in a continuum of multiparticle states, does not show up with a small 
      lattice volume.   The lattice spectrum is expected to become more realistic 
      with multiparticle operators included in the analyses and volume larger than the present value of $(2 \rm{fm})^3$.
      
      In Fig.~\ref{fig:J_spectrum_bands}, bands of states are 
      indicated by the shaded regions.  The lowest band has negative parity states, the next higher band
      has positive parity states, and as the mass increases, the bands alternate from one
      parity to the other in a staircase fashion.  It has been suggested that a restoration of chiral symmetry might
       occur at high excitation energy in QCD. ~\cite{Glozman:1999}
      That would show up as close to identical masses of higher states in both 
      positive- and negative-parity spectra.  
      The lattice spectra at $m_{\pi} = $396 MeV do not provide support for such a parity doubling.     

\section{$N^*$ states with the quantum numbers of $SU(6)\times O(3)$ states }

\begin{figure*}
\vspace{-.3in} \hspace{0.7in}\ig[width=0.75\tw]{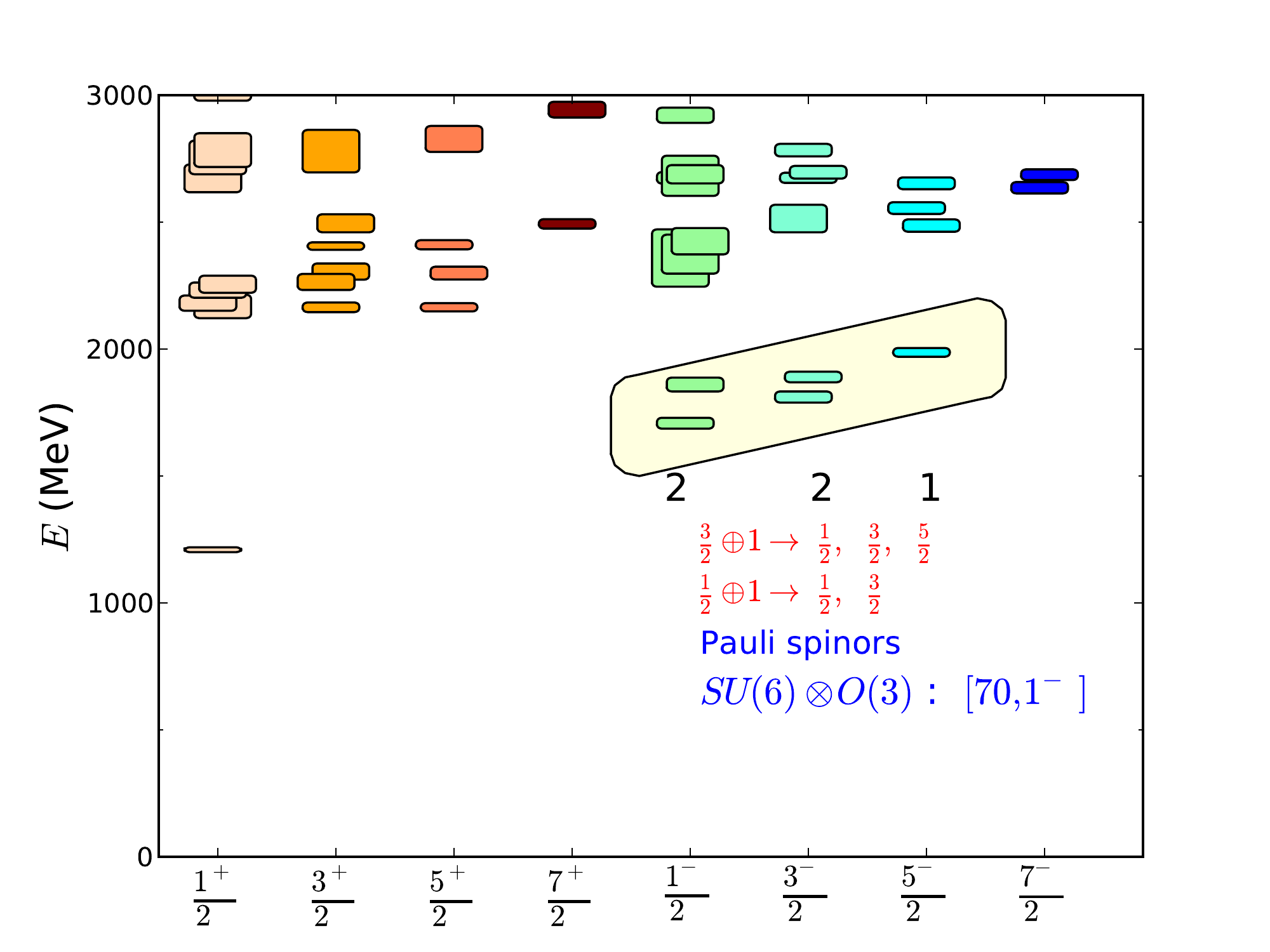}
\caption{Five $N^*$ states corresponding to the $[70, 1^-]$ multiplet of $SU(6)\times O(3)$ are
shown in the shaded area.
The $J$ values and parity are the same as can be formed by $S+ L \rightarrow J$, 
as tabulated in red, providing the same numbers of states of each spin
 as in the lattice spectrum.%, when orbital angular momentum and parity
% $L=1^-$ is coupled to the 
%possible spins of the three quarks ($S = \frac{1}{2}$ or $\frac{3}{2}$).  The resulting $J$ values
%give the same number of states of each spin as in the lattice spectrum.
 \label{fig:J_spectrum_band1}}
 \end{figure*}

      Focusing on the lowest band of $N^*$ states, Fig.~\ref{fig:J_spectrum_band1} 
      shows that the five states have spins based on the combinations of $S=\frac{1}{2}$ and 
      $S= \frac{3}{2}$ with orbital angular momentum and parity $L=1^-$.  The number of states is tabulated 
      inside the shaded region for each $J$ value and below that we tabulate the 
      construction of spins as $S \oplus L \rightarrow J$.
      These 5 states belong to the $[70, 1^-]$ multiplet of  $SU(6)\times O(3)$. Although
      the nucleon operators used are based on four-component Dirac spinors for each quark, the 
      lowest-energy 
      states are created predominantly by operators with only upper components of the Dirac spinors, i.e., Pauli spinors.
      
\begin{figure*}
\hspace{.7in}\ig[width=0.75\tw]{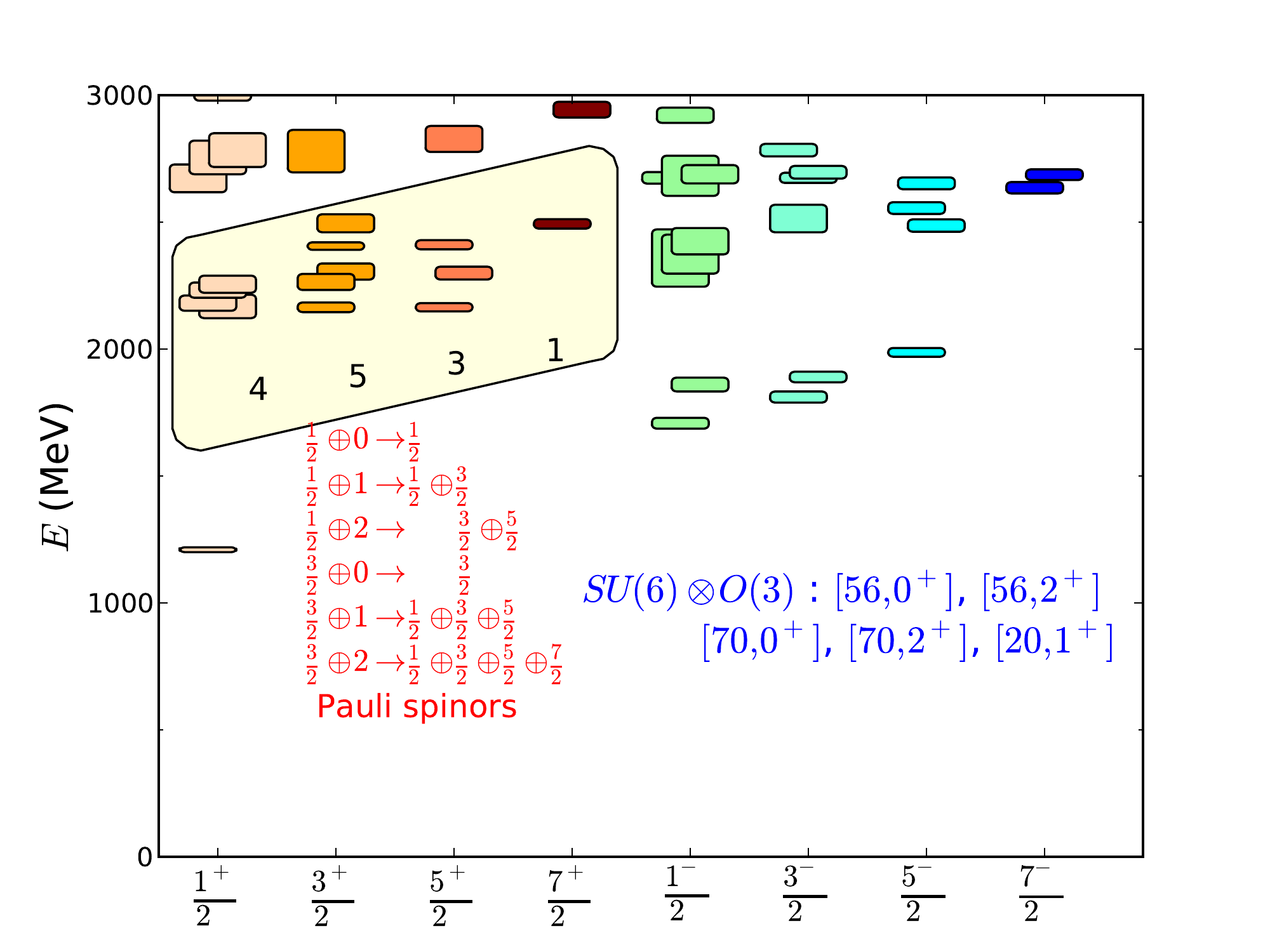}
\caption{Thirteen $N^*$ states corresponding to the $[56,0^+]$, $[56,2^+]$ $[70,0^+]$ $[70,0^+]$ and $[20,1^+]$
 multiplets of $SU(6)\times O(3)$ are shown in the shaded area. 
 The $J$ values and parity are the same as can be formed 
 by $S + L \rightarrow J$, as tabulated in red, when orbital angular momenta and parity 
 $L= 0^+, \ 1^+,\ $or $2^+$ are added to the possible quark spins, providing
 the same numbers of states of each spin
 as in the lattice spectrum. \label{fig:J_spectrum_band2}}
 \end{figure*}

      The next band at higher mass and positive parity contains 13 $N^*$ states in the shaded region of Fig.~\ref{fig:J_spectrum_band2}, with the number of states listed for each $J$ value.  Below that
      is a tabulation showing the $S \oplus L \rightarrow J$ constructions of the 13 states.  
      The operators used allow the construction to be determined in the same fashion as the spin, 
      i.e., by the overlaps.  
      As can be seen in the tabulation, the 13 states correspond to the positive parity states that can be made by combining S = $\frac{1}{2}$
      and $\frac{3}{2}$ with $L = 0^+, 1^+$ and $2^+$.  Nonzero L values are incorporated by covariant derivatives in the
      operators.  If the derivatives were omitted, only one $J= \frac{1}{2}^+$ state would appear in
      the shaded region rather than
      the 4 states that are obtained with derivatives.   That is interesting because previous lattice analyses that omit
       derivative operators generally have found one or two close-together excited $\frac{1}{2}^+$ states rather than 4 
       close-together states, as are obtained here.  Our results suggest that the Roper resonance 
       could have a more complex structure that would not be seen without the use of derivative operators. 
       It is an interesting question whether this pattern will still be evident when multiparticle 
       operators are included in the analysis.
       
       It has been suggested that a quark-diquark model might account for the $N^*$ spectrum. 
       The lattice results show that the $[20, 1^+]$ multiplet is realized, i.e., it gives the 5
       $L = 1^+$ constructions seen in the tabulations in Fig.~\ref{fig:J_spectrum_band2}.  Those states 
       are not consistent with a quark-diquark spectrum.   
    
      %%%%%%%%%%%%%%%
      \section{Summary}
      %%%%%%%%%%%%%%%
      The identification of spins of lattice states is significant because the usual method gives ambiguous 
      results, which is a consequence of the high degree of degeneracy in the spectra and uncertainties 
      in the determinations of lattice energies.
       The use of operators subduced from continuum spins
       allows robust spin identification. 
       That, in turn, allows 
      the interpretation of low-lying states in the lattice spectra in terms of the quantum numbers
      of $SU(6)\otimes O(3)$ symmetry.  The clarity of the spectrum with spins identified 
      provides contraindications with regard to three oft-discussed issues: parity doubling, quark-diquark models and
      a simple structure of the $N^* \frac{1}{2}^+$ (Roper) resonance.

       Support from U.S. Department of Energy contract DE-FG02-93ER-40762 is acknowledged.
       Computations were performed using {\tt Chroma}~\cite{Edwards:2004sx} 
       and {\tt QUDA}~\cite{Clark:2009wm,Babich:2010mu} at Jefferson Laboratory 
       under the USQCD Initiative and LQCD ARRA project.
            
\end{document}